\documentclass[final,5p,times,twocolumn]{elsarticle}

\usepackage{amssymb}
\journal{Elsevier Science}
\begin{document}
\begin{frontmatter}



\title{Cooperation in the snowdrift game on directed small-world networks
under self-questioning and noisy conditions}

\author[SE]{Tian Qiu \corref{cor1}}
\cortext[cor1]{School of Information Engineering, Nanchang Hangkong
University, Nanchang, 330063, China}
\ead{tianqiu.edu@gmail.com}

\author[SA]{Tarik Hadzibeganovic}
\author[SE]{Guang Chen}
\author[SB]{Li-Xin Zhong}
\author[SE]{Xiao-Run Wu}

\address[SE]{School of Information Engineering, Nanchang Hangkong University, Nanchang, 330063, China}
\address[SA]{Cognitive Science Section, Department of Psychology, University of Graz, 8010 Graz, Austria}
\address[SB]{School of Journalism, Hangzhou Dianzi University, Hangzhou, 310018, China}

\begin{abstract}
Cooperation in the evolutionary snowdrift game with a
self-questioning updating mechanism is studied on annealed and
quenched small-world networks with directed couplings. Around the
payoff parameter value $r=0.5$, we find a size-invariant symmetrical
cooperation effect. While generally suppressing cooperation for
$r>0.5$ payoffs, rewired networks facilitated cooperative behavior
for $r<0.5$. Fair amounts of noise were found to break the observed
symmetry and further weaken cooperation at relatively large values
of $r$. However, in the absence of noise, the self-questioning
mechanism recovers symmetrical behavior and elevates altruism even
under large-reward conditions. Our results suggest that an updating
mechanism of this type is necessary to stabilize cooperation in a
spatially structured environment which is otherwise detrimental to
cooperative behavior, especially at high cost-to-benefit ratios.
Additionally, we employ component and local stability analyses to
better understand the nature of the manifested dynamics.

\end{abstract}


\end{frontmatter}

% \linenumbers

\section{Introduction}
\label{1} Understanding the ubiquity of cooperative behavior in
complex biological, technological, social, and economic networks has
become one of the central research topics in the past years in a
variety of disciplines, including physics \cite {now10, srp06,
cfl09, swb09}. In the course of this challenging process of
scientific enquiry, the evolutionary game theory emerged as a common
mathematical framework for investigating how and why individuals
happen to overcome selfish behavior in order to help others and to
contribute to the common good \cite {axe81, axe84, hau04, now92a,
lie05, oht06, ant06, san08, gor09}.

Within this framework, two widely discussed games are the Prisoner's
Dilemma (PD) and the Snowdrift (SD) Game, where in both cases, the
bilateral cooperation yields the highest collective payoff, equally
distributed between the interacting players \cite {axe84}. However,
providing help to a neighboring individual is challenged by the
presence of an alternative defecting strategy, that warrants the
defector a higher benefit at the expense of a cooperating agent.

The major difference between the PD and SD games is manifested in
the actual ranking of the payoff values, i.e. in the manner by which
defecting agents are punished when facing one another. In the PD
game, a defecting agent facing another defector earns more than a
cooperator, while in the SD game, which can be seen as a modified PD
game, mutual defection gives the lowest payoff. We therefore have
the reversed situation where a cooperator obtains a higher payoff
than a defector when the latter interacts with another defecting
agent.

Thus, while in the PD game we have $T > R > P > S$ ranking of the
payoffs, for the SD game this ranking is given by $T > R > S > P$.
Here, the payoff for bilateral cooperation is denoted with $R$,
whereas $P$ represents the payoff for mutual defection. $T$ is the
temptation to defect by one of the two interacting sides, resulting
in the payoff $S$ (the "sucker's payoff") for the other, cooperating
player.

At first, this apparently minor difference between the two payoff
rankings might seem as rather unimportant, however, it has been
demonstrated \cite {hau04, now92a} that it can have a fundamental
impact on the final evolutionary outcome of a studied system in the
sense of altering the actual level of cooperative behavior whenever
interacting individuals are facing a social dilemma.

Whereas investigations on simple regular lattices yielded far more
converging results, studies of PD and SD games on complex networks
are still rather inconclusive about the overall structural effects
on cooperation \cite {now10, oht06}. Earlier reports suggested that
spatial structure generally tends to facilitate cooperative behavior
\cite {now92a, doe98, kil99}. Indeed, the scale-free topology \cite
{aba02} was found to be consistently beneficial for overcoming
selfish behavior in both PD and SD games \cite {san05}. However,
there have been many opposite findings in recent years in other than
scale-free topologies \cite {hau04, hau06, cas07, tpp07, abr01}. In
addition, a number of special conditions for the elevation and
inhibition of cooperation in both game types have been identified
\cite {shg06, fu07}.

Prompted by the observation that structured SD and PD games often
result in different densities of cooperative agents under different
connection topologies \cite {hau04, now92a, doe98, kil99, now93}, an
ever increasing effort has been dedicated to a better understanding
of the mechanisms involved in this 'spatial selection' process \cite
{oht06,ant06,san05,pac06,fu09,pen09,roc06,wu09,san06,zho06,wan06b,wan08b}.
Representative examples are the pertaining studies of the emergence
and survival of cooperation on directed networks \cite
{san05,kim02,jia08c,lim09b}. Directed interaction
\cite{zhu04,son09,pal07,fer10}, found in many social and economic
systems, implies here that an agent influencing its nearest-neighbor
may not be influenced by this neighbor in return. Thus, the
influence spreads in one direction only.

Besides the effects of topology, it has been demonstrated earlier
that the peculiarities of strategy update rules in theoretical
social dilemma models can play a decisive role in shaping the
evolutionary dynamics of cooperation \cite {tra10, wep09, qiu08a}.
Different strategy updating rules have been studied so far, for
example, the widely applied deterministic rule proposed by Nowak and
May \cite {now92a, now93}, and the stochastic evolutionary rule by
Szab\' o and T\" oke \cite {sza98}.

Motivated by the yet understudied effects of directed network
topologies and alternative updating rules, the scope of the present
paper is to identify the pertinent mechanisms leading to cooperative
behavior in SD-type of interactions on annealed and quenched
small-world networks \cite {znt02, vic04} with directed links. A
special attention is given to an alternative strategy updating rule,
the so-called self-questioning mechanism \cite {wan06b, gao07}, and
its influence on cooperative behavior under noisy conditions.

In the course of self-questioning, players typically compare their
current and the opposite strategies, and then eventually adopt a
more advantageous strategy with a certain probability. The
self-questioning mechanism can therefore be seen as a kind of a
self-evaluation procedure that helps in maximizing the benefits of
interacting agents.

To prevent deterministic behavior of the model and allow for
irrational decision making, we study the effects of noise,
implemented here as a low probability with which an agent shifts
from its own strategy to the strategy of a randomly chosen neighbor.
Considering such noise effects in models of social dynamics \cite
{wax95, lim06, had09, hyu10} might be useful for a better
understanding of the influence of bounded rationality in
evolutionary snowdrift and other games \cite {nix09}; see also Ref.
\cite {hel09} for cooperation under noisy conditions in the PD game
and Ref. \cite {duc09} for noise effects in both PD and SD games.
Finally, we employ an extensive local stability analysis inspired by
the Ref. \cite{wan06b}, and also implement a component analysis, in
order to better understand the nature of the observed cooperative
behavior.

The remainder of this paper is organized as follows. In the next
section, we detail the model structure and the associated game
dynamics. In Section 3, we outline the simulation results and
investigate the possible origin of the manifested cooperative
behavior. In Section 4, we analyze the influence of the introduced
strategy updating rule on cooperation under different reward and
noisy conditions. Section 5 describes analytic methods useful for a
further comprehension of the obtained results. Finally, we conclude
and suggest further
research directions in Section 6.\\

\section{Model}
\label{2}

In the present paper, all simulations are carried out for the
snowdrift (SD) game which we first outline below and then describe
the simulation setup.

In the SD game \cite {sug86}, players obtain a payoff $R=b-c/2$ when
they both cooperate, and a payoff $P=0$ when they both defect. When
one player cooperates but the other one defects, the defector gets
the payoff $T=b$, and the cooperator receives the payoff $S=b-c$.
The SD game refers to the case $b > c > 0$, leading to the payoff
ranking $T > R > S > P$. For simplicity, we assign $R=1$ to
characterize all payoffs by a single parameter $r = c/2 = c/(2b-c)$,
defined as the cost-to-benefit ratio. The payoffs are then rewritten
as $T = 1+r$, $R = 1$, $S = 1 - r$, and $P = 0$, or, when shown in a
rescaled payoff matrix

\begin{equation}
\begin{array}{c}
  \\
C \\
D
\end{array}
\begin{array}{c}
\, C \;\;\; D \\
\left( \begin{array}{cc}
1 & 1-r \\
1+r & 0
\end{array} \right)
\end{array}
\end{equation}

\noindent where $0 < r < 1$. Each matrix element represents here the
payoff of a player selecting a strategy from the left hand column
when the opponent selects a strategy listed in the top row.

In the initial configuration of our model, all players occupying the
network sites are assigned a strategy, either as a cooperator (C) or
as a defector (D), with equal probability. Thus, we assume that each
vertex of a network, ranging from a regular $L \times L$ square
lattice to a random graph by modifying $p$, is populated by a player
choosing just cooperation or defection.

Following the initial distribution of strategies, player
interactions and strategy updates are repeated in an iterative
fashion in accordance with the Monte Carlo (MC) simulation method.
Size-wise, simulations are conducted on networks with $L=100$,
$L=300$, and $L=500$. Furthermore, we separately investigate the
model behavior evolving on quenched and annealed small-world
networks. More specifically, the underlying evolutionary dynamics
are generated by the following elementary events:

(1) We first construct a small-world network with directed couplings
\cite{san02}. Starting with an $L \times L$ regular lattice, where
each node is bidirectionally connected with its four neighbors, we
rewire with probability $p$ each of the outward links to another
randomly chosen node.

(2) At each time step, players occupying the nodes of a directed
small-world network of a size $L^{2}$ obtain payoffs on the basis of
the above described payoff matrix by interacting with all
outward-connected neighbors in accordance with the SD game rules. A
player $i$ obtains a payoff $\overline{V_{i}} = V_{i}/4$ by
interacting with its four outward-linked neighbours, where $V_{i}$
is obtained by summing up the payoffs of all outward-linked
neighbors. Subsequently, by means of self-questioning in a virtual
game \cite {wan06b, gao07}, each player adopts its anti-strategy and
estimates a virtual payoff $\overline{V_{v}}=V_{v}/4$, where $V_{v}$
is obtained by summing up the virtual payoffs of all the
outward-linked neighbors. By comparing the virtual payoff with the
real payoff, players can estimate their optimal strategy
corresponding to the highest payoff and realize whether their
current strategy is of any advantage \cite {gao07}.

Specifically, player $i$ will shift from its current strategy to its
anti-strategy with the strategy transition probability $w$, defined
as the normalized difference between the estimated virtual payoff
$\overline{V_{v}}$ and the actual real payoff  $\overline{V_{i}}$:

\begin{equation}
w=\frac{\overline{V_{v}}-\overline{V_{i}}}{1+r} \label{eq.2}
\end{equation}

\noindent where $1+r$ in the denominator is only a normalization
factor \cite{zho06}, and as such, does not affect the results
reported in the subsequent sections. If $w>0$, then the current
strategy of player $i$ is updated with probability $w$, i.e.
replaced by the more advantageous anti-strategy; otherwise, it
remains unchanged.

(3) Network nodes are randomly scanned until each one is selected
once and then the rewiring process of step (1) and the strategy
evolution of step (2) are repeated.

Given these 3 outlined steps, we stress here that we investigate a
total of two different cases, i.e., 1) simulations on quenched and
2) annealed small-world networks.

In the first case, we assume that after a directed small-world
network is once established, the system evolves based on iterating
the step (2). Here the SD game dynamically evolves on the static
topology of the network with a fixed link directedness, which we
here call the quenched network \cite {znt02, sun09}.

In the second case, we consider the system evolution based on
repeating all outlined steps, i.e. the steps (1), (2) and (3). Thus,
after initially constructing a directed small-world network and then
following the update of strategies in accordance with step (2), all
nodes of the network are subsequently scanned and the rewiring
process of step (1) is re-initiated, leading thereby to a different
realization of rewired links. The process is then further cycled
with steps (2) and (3) following the step (1). In this particular
case, the SD game is said to evolve on an annealed network \cite
{zhu04, znt02, vic04}.

Finally, we note here again that in the SD game, the best action for
a player is to defect if the opponent cooperates, but otherwise it
is better to opt for cooperation. However, larger values of the
payoff parameter $r$ in the SD game generally tend to encourage
defection \cite {hau04, wan06b}. We therefore investigate what is
the critical value of $r$ for and under which cooperation is still
promoted given the specified structural and strategy update
characteristics of our model. Thus, it is important to investigate
how the frequency of cooperators $f_{c}$ dynamically relies on the
payoff parameter $r$ and on the spatial structure, i.e. the rewiring
probability $p$ of the network (where $f_{c}$ is the ratio of the
number of cooperators and the total number of
individuals in a population). \\

\section{Evolution of cooperation as a function of the spatial
structure ($f_{c}$ vs. $p$)} \label{3}

\subsection{$f_{c}$ vs. $p$ on quenched and annealed networks}

In Fig. 1(a) and (b), we show $f_{c}$ as a function of $p$ for
quenched and annealed networks with directed couplings. We
observe a symmetrical cooperation behavior around the payoff
parameter value $r=0.5$ for both networks. Cooperation dominates
for $r<0.5$, but is significantly suppressed for $r>0.5$ as $p$
increases. For $r=0.5$, $f_{c}$ remains stable for a wide range
of values of $p$. The displayed results in Fig. 1 are the outcomes of
simulations conducted on networks with $100 \times 100$ nodes.
Simulations on larger networks ($300 \times 300$ and $500 \times 500$)
did not yield significantly different results (not shown).

\begin{figure}[htb]
\centering
\includegraphics[width=8cm,height=5cm]{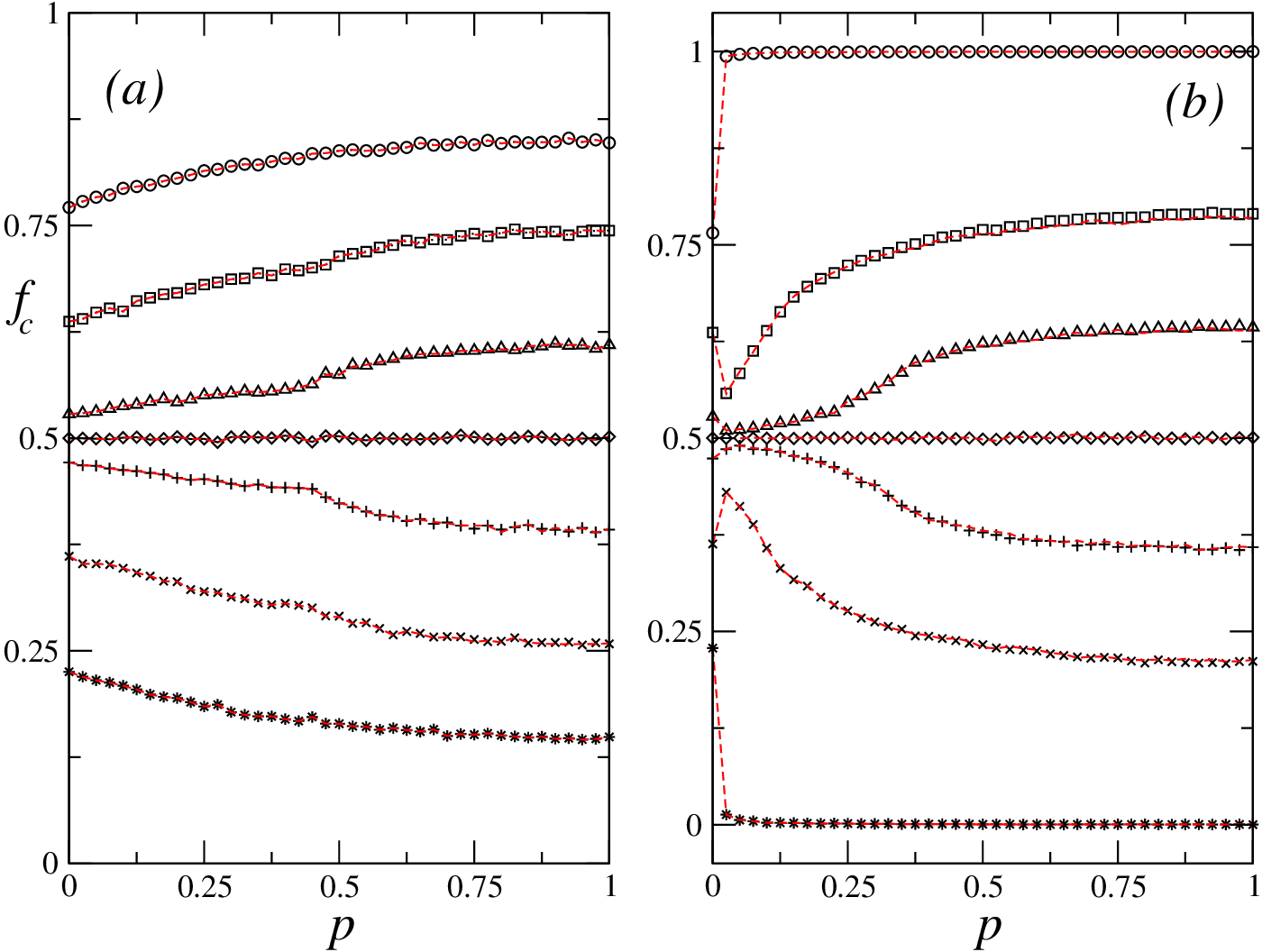}
\caption{\label{Fig:1} The frequency of cooperators $f_{c}$ as a
function of the rewiring probability $p$ is displayed in (a) and
(b), with circles, squares, triangles, diamonds, pluses, crosses and
stars standing for (from top to bottom) $r$=0.0, 0.1, 0.3, 0.5, 0.7,
0.9 and 1.0. (a) shows the quenched network and (b) the annealed
network. The red dashed curves are the theoretical fits.}
\end{figure}

For an intuitive understanding of the evolutionary behavior of
the system, we investigated the spatial patterns associated with
different values of $p$ and $r$. In Fig. 2, we illustrate the typical
patterns emerging on a quenched small-world network with directed links.
The annealed network shows similar behavior (here not shown). The
displayed patterns are independent of the initial configuration and
statistically static after the system reaches the steady state.
In Fig. 2(a), cooperating and defecting agents are scattered over
the lattice in a chessboard-like fashion for the parameter values $p=0.2$
and $r=0.2$, i.e. sites with one strategy are roughly regularly
surrounded by the sites with the opposite strategy.

We then increase the rewiring probability up to $p=0.8$ while keeping
the payoff value $r=0.2$ unchanged (Fig. 2(b)). It can be observed here
that at fixed and low $r$, cooperators start to form clusters as the
rewiring probability increases. In contrast, when increasing
and fixing the payoff parameter value at $r=0.8$, and again modifying
$p$ from $p=0.2$ to $p=0.8$, the observable strategy pattern transforms
from a state of scattered cooperators and defectors into the state of
clustered defectors (Fig. 2(c) and (d)). This is consistent with previous
findings showing that low $r$ stimulates cooperative behavior, whereas
elevated $r$ values tend to hinder the evolution of
cooperation \cite {hau04, wan06b}.

\begin{figure}[htb]
\centering
\includegraphics[width=8cm]{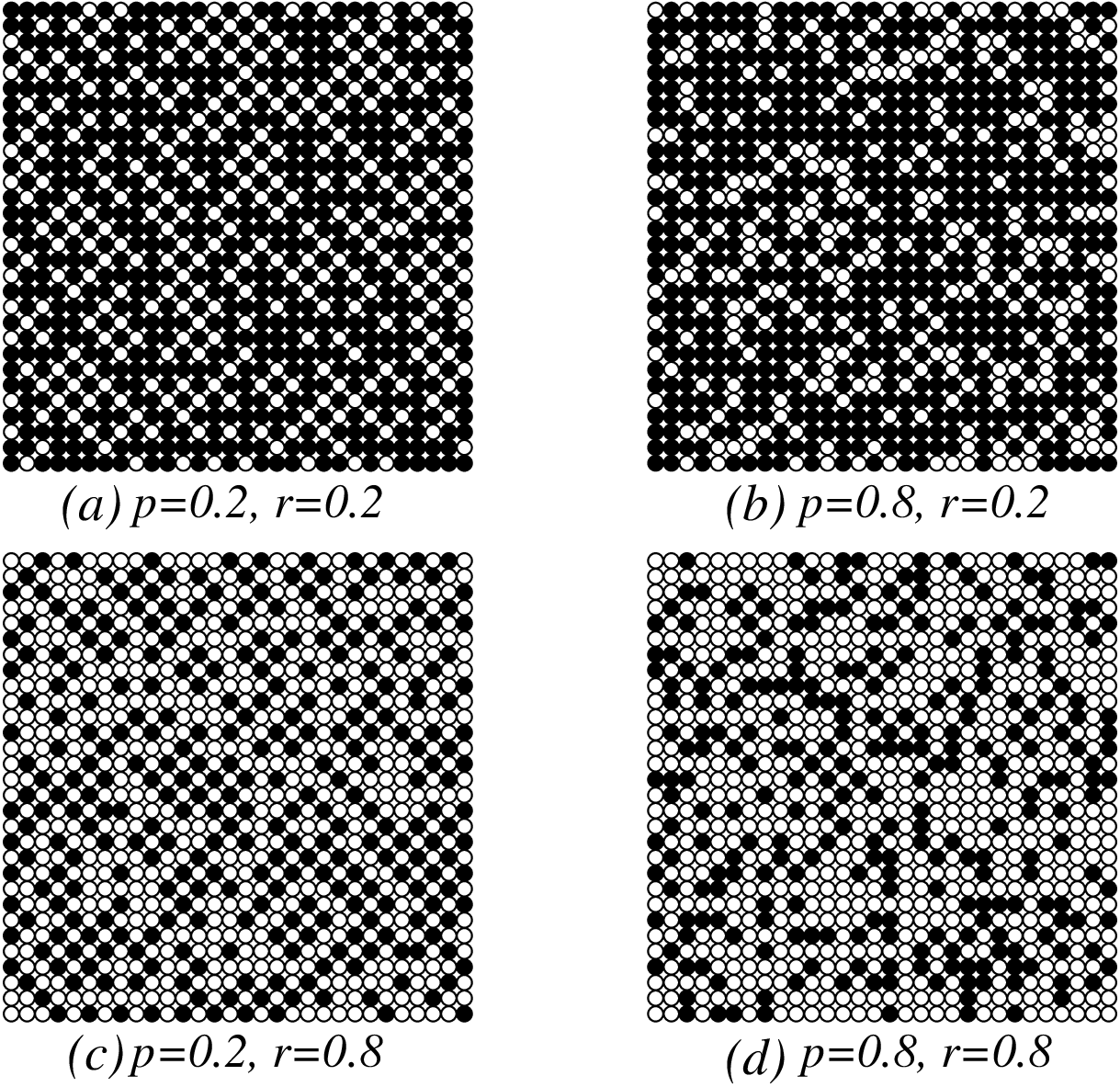}
\caption{\label{Fig:2} Representative simulation snapshots for the
spatial snowdrift game after a sufficiently long transient time. The
results are shown for a $30 \times 30$ portion of the full $100
\times 100$ lattice with typical spatial patterns emerging via agent
interactions on the quenched network, with C(D) nodes displayed as
solid(open) circles. (a) is for $p=0.2$ and $r=0.2$, (b) for $p=0.8$
and $r=0.2$, (c) for $p=0.2$ and $r=0.8$, and (d) for $p=0.8$ and
$r=0.8$.}
\end{figure}

\subsection{The effects of inward-link heterogeneity and long-range interaction}

To clarify whether the inward-link heterogeneity influences
cooperation, one can remove it by artificially fixing the number of
the inward links $k_{in}$. If cooperative behavior remains
qualitatively unchanged after such a modification, one can then
eliminate the possible influence of inward-links. Taking again the
quenched network as an example, in Fig. 3(a) we show $f_{c}$ as a
function of $p$ with fixed inward-links as $k_{in}=4$, which corresponds
to a regular lattice. We see that $f_{c}$ again shows a familiar
symmetrical behavior, as without this link fixation. Moreover, we
find that $f_{c}$ remains nearly steady for many different values of
$k_{in}$, as shown in Fig. 3(b). Additionally, we found that also the
annealed network simulations with fixed inward-links resulted in a
qualitatively similar behavior.

\begin{figure}[htb]
\centering
\includegraphics[width=8cm,height=5cm]{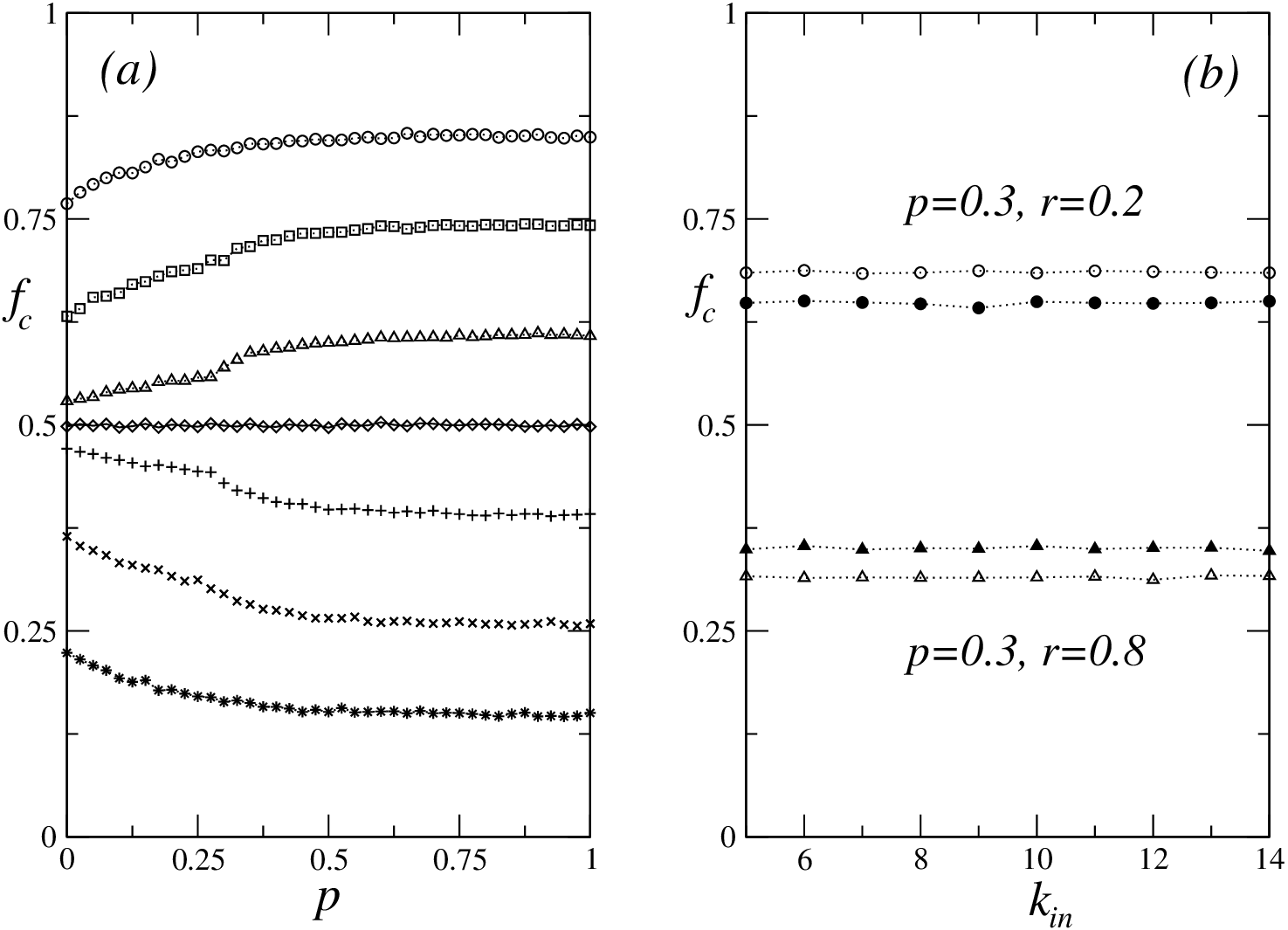}
\caption{\label{Fig:3} (a) The frequency of cooperators $f_{c}$ as a
function of the rewiring probability $p$ is displayed for the
quenched network with a fixed number of inward-links $k_{in}=4$.
Circles, squares, triangles, diamonds, pluses, crosses and stars
correspond to the payoff parameter values (from top to bottom)
$r$=0.0, 0.1, 0.3, 0.5, 0.7, 0.9 and 1.0. (b) $f_{c}$ as a function
of $k_{in}$ for a rewired probability $p=0.3$, and with open (solid)
circles standing for $r=0.2$ and open (solid) triangles for $r=0.8$,
both simulated on quenched (annealed) networks, respectively.}
\end{figure}

To further understand the origin of the symmetrical cooperation
observed in Fig.1, we study the possible effect of long-range
interactions by analyzing the model evolution on quenched and
annealed small-world networks with undirected links. As shown in
Fig. 4(a), the rewiring process has little effect on the cooperation
level for the undirected quenched network. However, the previously
discovered symmetrical behavior remains robust in this undirected
couplings version, under both quenched and annealed network regimes.

Having eliminated the long-range interaction and the inward-link
heterogeneity effects on the symmetrical behavior, we ask whether it
is the nature of the implemented strategy updating rule (generalized
self-questioning mechanism) that is actually causing the observed
cooperation symmetry. We address this issue in a more detail in the
subsequent section.

\begin{figure}[htb]
\centering
\includegraphics[width=8cm,height=5cm]{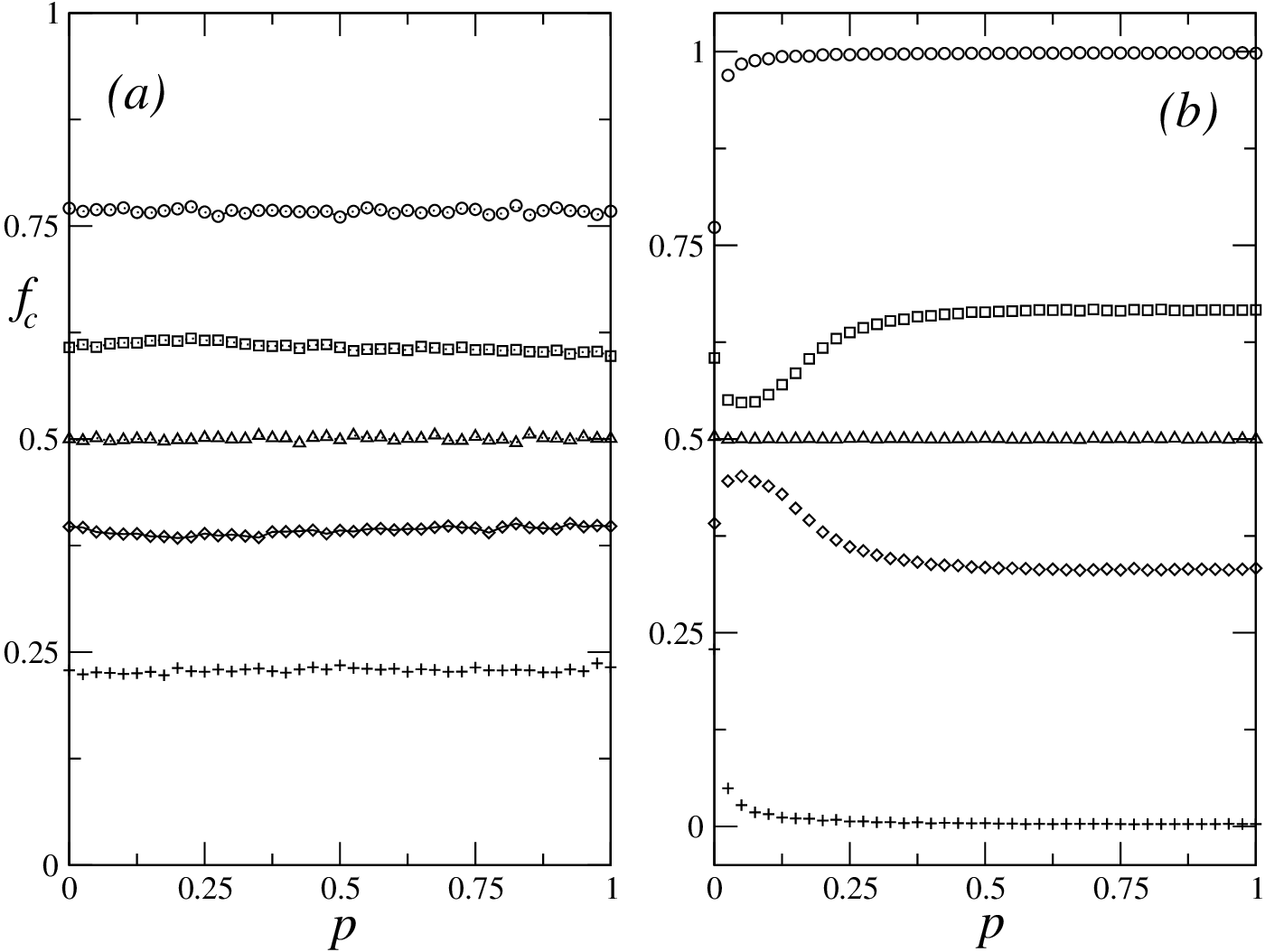}
\caption{\label{Fig:4} The frequency of cooperators $f_{c}$ as a
function of the rewiring probability $p$ in (a) quenched and (b)
annealed {\it undirected} small-world networks. Circles, squares,
triangles, diamonds and pluses correspond to (from top to bottom)
$r$=0.0, 0.25, 0.50, 0.75, and 1.00.}
\end{figure}

\subsection{The effect of noise}

To prevent the purely deterministic behavior and to enable
irrational choices, we introduced noise $ w_{noise} $ in the model
while keeping the model topology unchanged, i.e., the noise effects
were studied on quenched and annealed small-world networks with
directed links. More specifically, randomness was implemented as a
low probability with which a player shifts from its own strategy to
the strategy of a randomly chosen neighbor. Surprisingly, we find
that only after this direct modification of the previously
implemented strategy update rule the results started to change (see
Fig. 5). In particular, the observed cooperation symmetry is broken
after directly influencing the strategy updating mechanism via
noise. These changes are visible already after adding a rather small
amount of noise $ w_{noise}=0.05 $, but a clearly more pronounced
effect is observed for $ w_{noise} \geq 0.1 $ (Fig. 5).

\begin{figure}[htb]
\centering
\includegraphics[width=8cm,height=5cm]{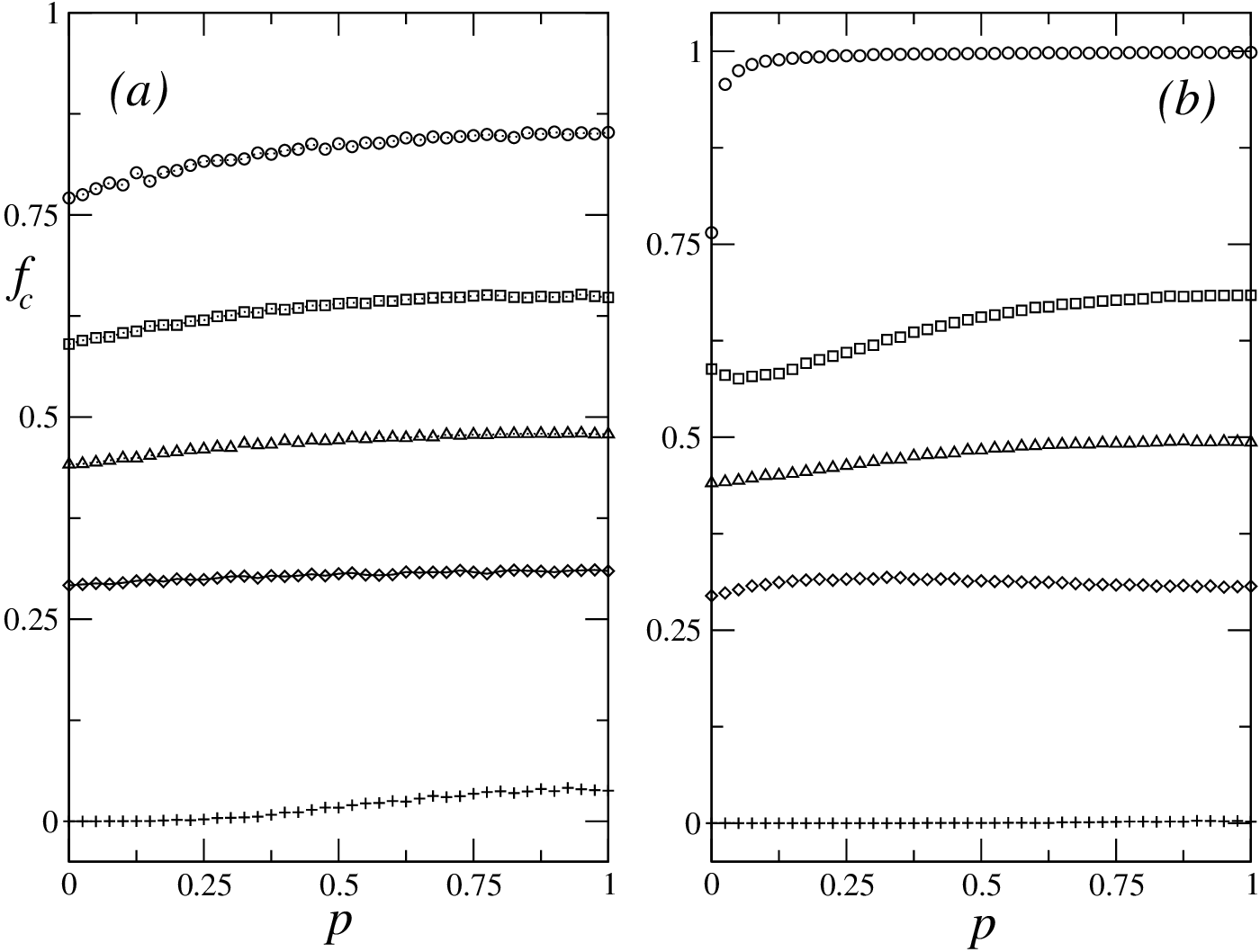}
\caption{\label{Fig:5} The effect of noise $w_{noise}=0.1$ on the
behavior of (a) quenched and (b) annealed small-world networks with
directed couplings. Circles, squares, triangles, diamonds and pluses
correspond to (from top to bottom) $r=0.0$, 0.25, 0.50, 0.75, and
1.00.}
\end{figure}

\begin{figure}[htb]
\centering
\includegraphics[width=8cm,height=5cm]{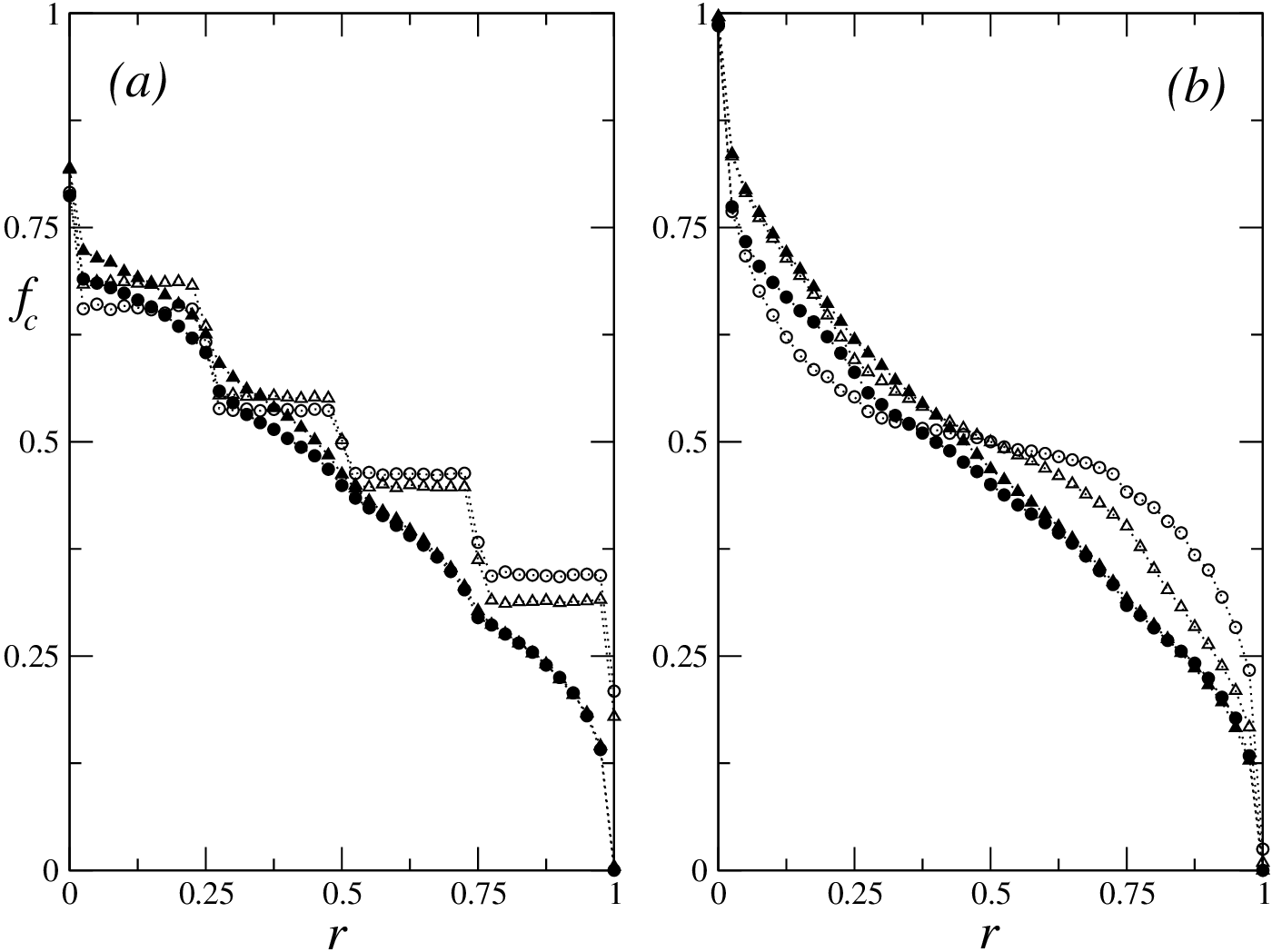}
\caption{\label{Fig:6} The fraction of cooperators $f_{c}$ as a
function of $r$ is displayed. (a) shows the results for the quenched
and (b) for the annealed network. The open circles and the triangles
correspond to simulation results with the self-questioning mechanism
without noise, and with the rewiring probabilities $p=0.1$ and
$p=0.3$, respectively. The solid circles and triangles correspond to
simulation outcomes obtained under noisy conditions, with the noise
parameter value set at $w_{noise}=0.1$ and the rewiring
probabilities $p=0.1$ and $p=0.3$, respectively.}
\end{figure}

Taken together, these results suggest that the symmetrical behavior
of $f_{c}$ arises due to the characteristics of the implemented game
rule, i.e., due to the so-called self-questioning mechanism \cite
{wan06b,gao07} which drives the process of strategy updating in our
model. In contrast, enabling irrational choice behavior via noise
breaks the cooperation symmetry, however, it is unclear yet whether
the self-questioning mechanism can support cooperative clusters
against the destructive effect of noise also under large-reward
conditions, i.e. at high cost-to-benefit ratios. This is
investigated in detail in the
following section. \\

\section{Dependence of $f_{c}$ on $r$}

\subsection{The self-questioning mechanism without noise}
\label{5}

We now investigate the fraction of cooperators $f_{c}$ as a
function of the payoff parameter $r$ in the SD-game on quenched
and annealed small-world networks with an implemented generalization
of the  self-questioning mechanism described in Section 2. Here
we care more about the evolution of cooperation at a relatively small
value of the rewiring probability $p$, because larger $p$ typically
results in too many random realizations of long-range interactions among
interacting players.

As shown in Fig. 6 (open circles and triangles), $f_{c}$ is observed
to decay with $r$ both in quenched and annealed networks with
directed couplings. More interestingly, step-wise decay of
cooperation not seen in the annealed network and sharp transitions
at the critical points $r_{c}=0.25$, 0.50 and 0.75 are observed for
the quenched small-world network.

The decay dependence of $f_{c}$ on $r$ can be explained analytically
from the payoff matrix shown in Table 1, where the total transition
probabilities $W_{C \rightarrow D}$ from C to D and $W_{D
\rightarrow C}$ from D to C can be calculated as
$4-\frac{11}{2(1+r)}$ and $-4+\frac{13}{2(1+r)}$; they are
monotonously increasing and decaying with $r$, respectively.

To understand the critical transition points $r_{c}$, we apply an
extensive local stability analysis, following the method detailed in
Ref. \cite{wan06b}. At each critical point $r_{c}$, the payoff of a
cooperating player should equal that of a defector. The local
stability equation is then written as

\begin{equation}
m+(k_{out}-m)(1-r_{c})=(1+r_{c})m, \label{eq.3}
\end{equation}

\noindent where $k_{out}$ is the number of the out-going links, with
$k_{out}=4$ in our model, and $m$ is the number of C neighbours. One
can get $r_{c}=(k_{out}-m)/k_{out}$, and therefore obtain $r_{c}$=
0.25, 0.50 and 0.75, which is in excellent agreement with the obtained
simulation results.

\subsection{Self-questioning under noisy conditions}

To further understand the nature of the observed cooperation in the
SD-game on complex networks with the self-questioning mechanism, we
investigate the evolution of $f_{c}$ as a function of $r$ when the
strategy updating is disturbed by noise, i.e. by the introduction of
random choices. As can be seen in Fig. 6 (solid circles and
triangles), the level of cooperation in the total population of
agents decreases as a function of $r$, also in the presence of
noise. Slight concavities of the solid circles and triangles are
visible as remnants of the previously observed critical points
$r_{c}$ in the quenched network, however, the previous step-wise
decay is now almost totally smoothed out by noise.

Furthermore, we see that a fair amount of noise generally hinders
the evolution of cooperation for a wide range of values of $r$
(especially for larger $r$), and this holds true for both quenched
and annealed networks. However, in the absence of noise, the process
of repeated self-questioning can re-establish symmetrical
cooperation behavior and elevate altruism among interacting agents
even under large values of $r$ (see the curves with open circles and
triangles in Fig. 6). These results suggest that self-questioning
can serve as a highly useful cooperation supporting mechanism in
noise-free environments, but is not robust enough against the
destructive effects of noise-driven random choices.

\section{Component Analysis}
\label{5}

To further understand the evolution of cooperation in our model, we
apply a theoretical analysis, which we here call the component
analysis. We first specify that each player
occupying a network vertex can have a total of five possible
neighborhood configurations. The payoff matrix can then be
analytically obtained from the strategy update rules, as shown in
Table~\ref{tab.1}. $E_{C}$ and $E_{D}$ are the payoffs of the
cooperating (C) and defecting (D) players, $w_{C \rightarrow D}$ is
the transition probability with which a player shifts from C to D,
and $w_{D \rightarrow C}$ is the the transition probability for
shifting from D to C. The frequency of cooperators $f_{c}$ is then
written as

\begin{equation}
f_{c}=\sum_{i=1}^{5}g_{C}^{i}w_{C \rightarrow C}^{i}+g_{D}^{i}w_{D
\rightarrow C}^{i}, \label{eq.4}
\end{equation}

\noindent where $i=1,...,5$ labels the corresponding neighbour
configurations shown in the leftmost column of the
Table~\ref{tab.1}. $g_{C}^{i}$ and $g_{D}^{i}$ are the proportions
of the C and D players in the system, with
$\sum_{i=1}^{5}(g_{C}^{i}+g_{D}^{i})=1$. $w_{C \rightarrow C}^{i}$
is the probability that C players do not change their strategy, with
$w_{C \rightarrow C}^{i}=1-w_{C \rightarrow D}^{i}$. Due to the
difficulty of obtaining the $g_{C}^{i}$ and $g_{D}^{i}$
analytically, we compute these proportions from numerical
simulations. As shown in Fig. 1(a) and (b) with the red dashed
curves, the theoretical fits of $f_{c}$ vs. $p$ are in a very good
agreement with those obtained from numerical simulations.

\begin{table}
\caption{The payoff matrix in the SD game with a generalized
self-questioning mechanism. From left to right, the columns
correspond to the neighborhood configuration, the payoff of the C
player $E_{C}$, the payoff of the D player $E_{D}$, the probability
of transition from C to D $w_{C \rightarrow D}$, and the probability
of transition from D to C $w_{D \rightarrow C}$, respectively.}
\label{tab.1}
\begin{center}
\begin{tabular}{|c|c|c|c|c|c|}
\hline & $E_{C}$ & $E_{D}$ & $w_{C \rightarrow D}$ & $w_{D \rightarrow C}$\\
\hline
C C C C & 1 & $1+r$ &  $1-\frac{1}{1+r}$ & $0$\\
C C C D & $1-\frac{r}{4}$ & $\frac{3}{4}+\frac{3r}{4}$ & $1-\frac{5}{4(1+r)}$ & $-1+\frac{5}{4(1+r)}$ \\
C C D D & $1-\frac{r}{2}$ & $\frac{1}{2}+\frac{r}{2}$ & $1-\frac{3}{2(1+r)}$ & $-1+\frac{3}{2(1+r)}$\\
C D D D & $1-\frac{3r}{4}$ & $\frac{1}{4}+\frac{r}{4}$ & $1-\frac{7}{4(1+r)}$  & $-1+\frac{7}{4(1+r)}$\\
D D D D & $1-r$ & 0 & $0$ & $-1+\frac{2}{1+r}$\\

\hline
\end{tabular}
\end{center}
\end{table}

\section{Conclusions}
\label{6}

In the present paper, we have studied the evolutionary SD-game with
a generalized self-questioning updating mechanism. The model was
investigated on both annealed and quenched small-world networks with
directed couplings. Agent-based computer simulations and
semi-analytic results have been presented.

We found that the observed symmetrical cooperation effect around the
payoff parameter value $r=0.5$ was size-invariant and independent
from both the inward-link heterogeneity and long-range interactions
in the studied small-world network. Moreover, our results suggest
that the self-questioning updating mechanism might be necessary to
stabilize cooperation in a spatially structured environment which is
otherwise not beneficial or even detrimental to cooperative
behavior. Our results thus support the previous findings showing
that spatial structure should not necessarily be seen as a general
promoter of cooperation \cite {hau04}, and that one should rather
avoid the conceptual flaws of many recent studies according to which
an implemented topology automatically enhances cooperative behavior
by default.

Compared with the previous work in this field \cite {wan06b, gao07},
our version of the self-questioning mechanism showed clear
advantages in (re)establishing and sustaining symmetrical
cooperative behavior. However, we found that the cooperation
promoting effects of the implemented updating rule were not robust
enough under noisy conditions. On the other hand, in the absence of
noise, the self-questioning mechanism was able to recover elevated
altruism even at large cost-to-benefit ratios. We therefore argue
that our updating mechanism can function as an efficient cooperation
elevator in complex environments where the impact of noise is
insignificant or totally absent.

In the presented model, the rewiring of the network was independent
from the actual cooperation status of the individual network nodes.
It would therefore be much more interesting in future studies to let
links between two rarely cooperating nodes vanish with higher
probability than links between frequently cooperating nodes \cite
{sta06}. In future investigations one should also employ the
simulated annealing method \cite {kir83}, in order to better avoid
the local extrema trap, and to enable the system to get into the
global extremum.

As suggested in Ref. \cite {kum09}, future computer simulations
should also explore networks with weighted links, since in the
present paper all network connections were equally strong. However,
in real social networks, people typically have many initial
connections, but after a sufficient time, only a few of them survive
as strong and important \cite {oht06,kar04}.

Another line of future generalizations that could potentially make
our model more realistic would include simulations with reproducing
agents \cite{lim09b}, different migratory behaviors of interacting
individuals \cite {had09,hel09}, bottleneck and ageing effects on
cooperation \cite{mal08}, or the influence of the approaching
extinction of a studied population \cite {mal07}.

In sum, presented results demonstrate that under noise-free
conditions, the self-questioning updating mechanism can become
advantageous for cooperation in the evolutionary snowdrift game on
both annealed and quenched small-world networks with directed links.
It remains therefore a challenge for future studies to search for
advanced modifications of the introduced updating rule which could
then fully resist the destructive effects of noise.

\bigskip
{\textbf{Acknowledgments:}}

We are grateful to Prof. Dayin Hua for fruitful discussions on a
variety of topics related to the present study. We are especially
thankful to Prof. Dietrich Stauffer for many useful suggestions and
comments on a previous version of the present paper. This work was
partly supported by the National Natural Science Foundation of China
(Grants No. 10805025, 10774080), Jiangxi Provincial Educational
Foundation of China under Grant No. GJJ08231, and Zhejiang Social
Sciences Association under Grant 08N51.

\bibliographystyle{elsarticle-num}
\bibliography{rint01}






% \end{thebibliography}

\end{document}